\documentclass[prl,twocolumn,superscriptaddress,floatfix,nopacs]{revtex4}
\usepackage{amsmath,amsfonts,epsf,dcolumn,natbib,graphicx}
\usepackage{ascii}
\usepackage{xcolor}
\usepackage{soul}
\usepackage{mathtools}
\usepackage{verbatim}

\newcommand{\be}{\begin{equation}}
\newcommand{\ee}{\end{equation}}
\newcommand{\beq}{\begin{equation}}
\newcommand{\eeq}{\end{equation}}
\newcommand{\bea}{\begin{eqnarray}}
\newcommand{\eea}{\end{eqnarray}}

\begin{document}
\title{ Non-empirical Semi-local Free-Energy Density Functional \\ for Matter 
Under Extreme Conditions }
\author{Valentin V.~Karasiev}
\email[Corresponding author. 
\\Electronic address: ]{vkarasev@lle.rochester.edu}
\affiliation{Quantum Theory Project, Department of Physics and Department 
of Chemistry, P.O.\ Box 118435, University of Florida, Gainesville FL 32611-8435, USA
}
\affiliation{Laboratory for Laser Energetics,
University of Rochester, 250 East River Road, Rochester NY 14623, USA}
\author{James W.~Dufty}
\affiliation{Department of Physics, P.O.\ Box 118435, University of Florida, Gainesville
FL 32611-8435, USA}
\author{S.B.~Trickey}
\affiliation{Quantum Theory Project, Department of Physics and Department of Chemistry,
P.O.\ Box 118435, University of Florida, Gainesville FL 32611-8435, USA}
\date{02 Nov.\ 2017} 

\begin{abstract}

\noindent Realizing the potential for predictive density functional calculations
of matter under extreme conditions depends crucially upon having an
exchange-correlation (XC) free energy functional accurate over a wide
range of state conditions. Unlike the ground-state case, no such
functional exists.  We remedy that with systematic construction of a
generalized gradient approximation XC free-energy functional based on
rigorous constraints, including the free energy gradient
expansion. The new functional provides the correct temperature
dependence in the slowly varying regime and the correct zero-T,
high-T, and homogeneous electron gas limits. 
Its accuracy in the warm dense matter regime is attested
by excellent agreement of the calculated deuterium equation of state
with reference path integral Monte Carlo results at intermediate and
elevated T.  Pressure shifts 
for hot electrons in  compressed static fcc Al
and for low density Al demonstrate the
combined magnitude of thermal and gradient effects handled well by this
functional over a wide T range.

\end{abstract}

\maketitle





\textit{Introduction}. Interest 
in high-energy density  
physics (HEDP) is burgeoning \cite{WitteEtAl2017,BenedictSurh...Murillo2017,DesjarlaisScullardBenedictWhitleyRedmer2017,DornheimGrothMaloneSchoof..BonitzPP2017,DriverMilitzer2017,DriverSoubiranZhangMilitzer17,FeldmanDyerKukDitmire17,HarbourDharmaWardanaKlugLewis17,HuGaoDingCollinsKress17,KnudsonDesjarlais17,Vaisseau...Santos2017,WitteShihabGlenzerRedmer17,ZhangDriverSoubiranMilitzer17,BurkeSmithGrabowskiPribramJones16,DanelKazandjianPrion2016,Knudson..Redmer.Science2015,BeckerEtAl2014,BenuzziMonaixEtAl2014,Ruter.Redmer.PRL2014,IPAMreview,CopariEtAl2013,WilsonMilitzer12,Drake10,Knudson.Desjarlais.PRL2009,HEDLPreport.2009}.  
Notable facilities include  
the Matter in Extreme Conditions instrument
at the Linac Coherent Light Source (LCLS), the ORION Laser, 
the OMEGA Laser System, the Sandia National Laboratories Z machine
and the GSI PHELIX-laser facility 
\cite{Fletcher.NP2015,LCLS.2015,LCLS.NC2016,ORION.PRL2013,OMEGA.PRL2015,OMEGA.PRB2017,Knudson..Redmer.Science2015,KnudsonDesjarlais17,PHELIX.EPL2016}.
A particularly challenging state-condition regime is so-called warm 
dense matter (WDM). Characterized by elevated temperature T and a
wide range of pressures $P$, best practice for predictive WDM/HEDP calculations
is to use finite-T density functional theory 
\cite{Mermin65,Stoitsov88,Knudson.Desjarlais.PRL2009,Ruter.Redmer.PRL2014,
  Knudson..Redmer.Science2015,IPAMreview,WitteEtAl2017,Vinko..Wark.NC2013,Hu.PRL2017} 
to drive {\it ab initio} molecular dynamics (AIMD) 
\cite%
{Barnett93,MarxHutter2000,Tse2002,MarxHutter2009}. 
Reliable predictions require accurate free-energy density functionals
adequate to the state conditions.   

Currently almost all AIMD matter-under-extreme-condition simulations
use a ground-state exchange-correlation (XC) functional.  Unlike the
ground-state situation, there are only a few very approximate
free-energy XC functionals.  Despite the fact that 
density-gradient dependence (via the 
generalized gradient approximation, GGA) is well-established as
{\it essential} reasonable ground state descriptions, there
is no counterpart GGA XC free energy functional.
The simplest XC free energy functional is the local density approximation
(LDA), based on the density and T dependencies of the
homogeneous electron gas (HEG) free energy. Our recent parametrization
of path-integral Monte Carlo data for the HEG at finite-T provides a
suitable LDA (the KSDT functional) \cite{KSDT,corrKSDT}.  As with the
ground state, the finite-T LDA is not enough for predictive purposes.
Ref.\ \onlinecite{FxcSignif} showed that accurate predictions require
an XC free-energy functional which incorporates both intrinsic T and
density gradient effects. Earlier thermal Hartree-Fock results \cite%
{PRE86,SjostromHarrisTrickey12} are consistent with that assessment.
Until now, the few XC free energy functionals that might meet
the need include random-phase approximation and classical mapping
functionals \cite{PDW84,PDW2000} and a combination (``SD14'')
\cite{Sjostrom.Daligault.2014A} of gradient-dependence from
ground-state GGA $E_{\mathrm{xc}}^{\mathrm{GGA}}$ and explicit
T-dependence only from the LDA XC free energy.  
 None of these is a 
true finite-T GGA, in stark contrast with 
the ground-state situation.  

In this Letter, we remedy that major deficiency by providing an authentic 
GGA XC free energy
density functional.  We 
describe constraints and limits for identifying suitable reduced
density gradient variables and constructing a proper,
non-empirical GGA XC free-energy functional. 
Analogously with  Ref.\ \onlinecite{KarasievSjostromTrickey12A} 
for construction of a non-interacting free-energy GGA, here we develop the 
generalization of 
$\mathrm{T}=$ 0K XC parametrization variables to $\mathrm{T} >$ 0K.  We 
construct new X and C enhancement factors that handle 
the unique properties of those variables correctly. We illustrate the 
efficacy and accuracy of
the new functional with a calculation on deuterium 
and two sets of calculations on Al. Significant deficiencies in the use 
of ground-state XC functionals are exposed. In particular, only the
new KDT16 functional gives agreement with PIMC results on deuterium.  
In Al, the $P$ shifts $\Delta P(T)$ between 
LDA and the ground state PBE functional \cite{PerdewBurkeErnzerhof96}  
have the wrong sign
compared to those from all known constraint-based free-energy XC functionals 
(new KDT16, SD14, KSDT).

\textit{Requisites}. Systematic construction of a GGA XC free energy rests on three requisites: 
(a) The finite-T LDA XC 
must be recovered 
in the HEG and high-T limits; 
high-T effects within the LDA XC free energy prevail 
over the 
gradient contributions there. 
(b) Proper T-dependent reduced density gradient variables
must be consistent with the XC free-energy gradient expansion. 
Use of those variables in the GGA enhancement factors for X and C must
recover 
the weakly inhomogeneous electron gas regime correctly. 
(c) As $\mathrm{T}\rightarrow 0$K, the GGA XC free-energy must reduce to a
ground-state functional which satisfies known constraints for the
ground-state XC energy. 

Regarding item (c), though there are more refined
non-empirical ground-state GGA XC functionals \cite{PW91like}, we
choose to recover the popular PBE functional \cite%
{PerdewBurkeErnzerhof96}. This choice enables use of existing
resources such as projector augmented wave (PAW) data sets and
pseudopotentials. The XC free-energy functional then is constructed by
adding finite-T constraints [according to requisites (a) and (b)] to ground-state ones used to determine PBE.

\textit{Finite-T Gradient Expansion}. 
As with the ground state, the second-order gradient correction
for the XC free-energy density is 
\cite{Hohenberg-Kohn,KohnSham,Geldart.CJP.I,Geldart.CJP.II,Geldart.CJP.III,Geldart.TCC.1996} 
\begin{equation}
nf_{\mathrm{xc}}^{(2)}(n,\nabla n,\mathrm{T})= {\tfrac{1}{2}} g_{\mathrm{xc}%
}^{(2)}(n,\mathrm{T})|\nabla n({\mathbf{r}})|^2 \, .  \label{gxc}
\end{equation}
Ref.\ \cite{Sjostrom.Daligault.2014A} provides $ g_{\mathrm{xc}%
}^{(2)}(n,\mathrm{T})$ numerically.  
In terms of the ground-state reduced density gradient
variable $s=|\nabla n|/2(3\pi^2)^{1/3}n^{4/3}$ and
 reduced temperature $t = {%
\mathrm{T}}/{\mathrm{T}}_F \equiv 2k_B\mathrm{T}/[3\pi^2n]^{2/3}=(2/3)^{2/3}
I_{1/2}^{-2/3}(\beta\mu)$, with ${\mathrm{T}}_F$ the Fermi temperature, the 
X and C contributions are 
\begin{eqnarray}
f_{\mathrm{xc}}^{(2)}(n,\nabla n,\mathrm{T})&=& C_{\mathrm{x}}^{(2)} {%
\varepsilon}_{\mathrm{x}}^{\mathrm{LDA}}(n)s^2(n,\nabla n) \widetilde B_{%
\mathrm{x}}(t)  \notag \\
&+& C_{\mathrm{c}}^{(2)} n^{1/3}s^2(n,\nabla n) \widetilde B_{\mathrm{c}%
}(n,t) \, .  \label{gxc2}
\end{eqnarray}
 Here 
$\beta\equiv 1/k_B{\mathrm{T}}$, 
$I_{k}$ is a Fermi-Dirac integral \cite{Blakemore1982,Bartel..Durand.1985},  $%
{\varepsilon}_{\mathrm{x}}^{\mathrm{LDA}}$ is the ground-state LDA exchange
energy per electron, and $\widetilde B_{\mathrm{x}}(t)$ is a combination of
Fermi-Dirac integrals (details below), hence a function of $t$ alone.  
Note the X gradient correction factorization into a 
product of the familiar $s^2$ and a function of $t$ alone. Note also 
that $\widetilde B_{\mathrm{c}}$ depends on both $n$ and $t$ \cite{CoeffComment1}. That differences causes the finite-T GGA
for X and C to be treated separately.  

\textit{Finite-T GGA exchange}. GGA functionals are defined with respect to
LDA. The X free-energy per particle LDA at  
chemical potential $\mu$  has the factorized
form \cite{Perrot.1979} 
\bea
f_{\mathrm{x}}^{\mathrm{LDA}}(n,\mathrm{T})&=& {\varepsilon}_{\mathrm{x}}^{%
\mathrm{LDA}}(n) \widetilde A_{\mathrm{x}}(t) \, ,  \label{fxLDA2} \\
\widetilde A_{\mathrm{x}}(t)&=&\frac{t^2}{2} \int_{-\infty}^{(\beta\mu)}
I_{-1/2}^{2}(\eta) d\eta \,.  \label{chi}
\eea
To exploit this form, the second-order gradient
expansion (GE2) for the X free energy (recall Eq.\ (\ref{gxc2}))
can be written \cite%
{Geldart.CJP.I,Geldart.CJP.II,Geldart.CJP.III,Geldart.SSC.1994,Geldart.TCC.1996,
CoeffComment2}
\begin{equation}
f_{\mathrm{x}}^{\mathrm{GE2}}(n,\nabla n,\mathrm{T})= f_{\mathrm{x}}^{%
\mathrm{LDA}}(n,\mathrm{T}) \Big(1+\frac{8}{81}\frac{\widetilde B_{\mathrm{x}%
}(t)}{\widetilde A_{\mathrm{x}}(t)} s^2(n,\nabla n)\Big) \, .  \label{fxSGA2}
\end{equation}
\begin{equation}
\widetilde B_{\mathrm{x}}(t) := \left(\frac{3}{2}\right)^{4/3}
I_{1/2}^{4/3}(\beta\mu) \Big[\Big(\frac{I_{-1/2}^{\prime }(\beta\mu)}{%
I_{-1/2}(\beta\mu)}\Big)^2 -3\frac{I_{-1/2}^{\prime \prime }(\beta\mu)}{%
I_{-1/2}(\beta\mu)}\Big]\, .  \label{Bx2}
\end{equation}
Primes indicate differentiation with respect to the argument. Details 
and accurate fits for $\widetilde A_{\mathrm{x}}$ and $\widetilde B_{%
\mathrm{x}}$ as explicit functions of $y\equiv 2/3t^{3/2}$ (or functions of $%
t$ after a variable change) are in Ref.\ \onlinecite{FDfits}.

GGA construction requires 
identifying  appropriate reduced gradient variables from
the gradient expansion.  Eq.\
(\ref{fxSGA2}) exposes the X free-energy appropriate reduced density
gradient as 
\begin{equation}
s_{2\mathrm{x}}(n,\nabla n, \mathrm{T})\equiv s^2(n,\nabla n) \frac{%
\widetilde B_{\mathrm{x}}(t)}{\widetilde A_{\mathrm{x}}(t)} \, .  \label{sx}
\end{equation}
(Remark: we cannot define the appropriate variable linearly 
in $s$ because $%
\widetilde{B}_{\mathrm{x}}(t)$ has both signs; see below.) Then the 
X free-energy GGA becomes 
\begin{equation}
{\mathcal{F}}_{\mathrm{x}}^{\mathrm{GGA}}[n,T]= \int n f_{\mathrm{x}}^{%
\mathrm{LDA}}(n,\mathrm{T}) F_{\mathrm{x}}(s_{2\mathrm{x}})d\mathbf{r}\; ,
\label{FxGGA}
\end{equation}
an evident generalization of the GE2 X free-energy,
Eq.\ (\ref{fxSGA2}). It is straightforward to show that 
\begin{equation}
\lim_{\mathrm{T} \rightarrow 0} s_{2\mathrm{x}}(n,\nabla n, \mathrm{T}%
)=s^2(n,\nabla n) \, .  \label{limschi}
\end{equation}

\begin{figure}
\includegraphics*[angle=-00,height=2.8cm]{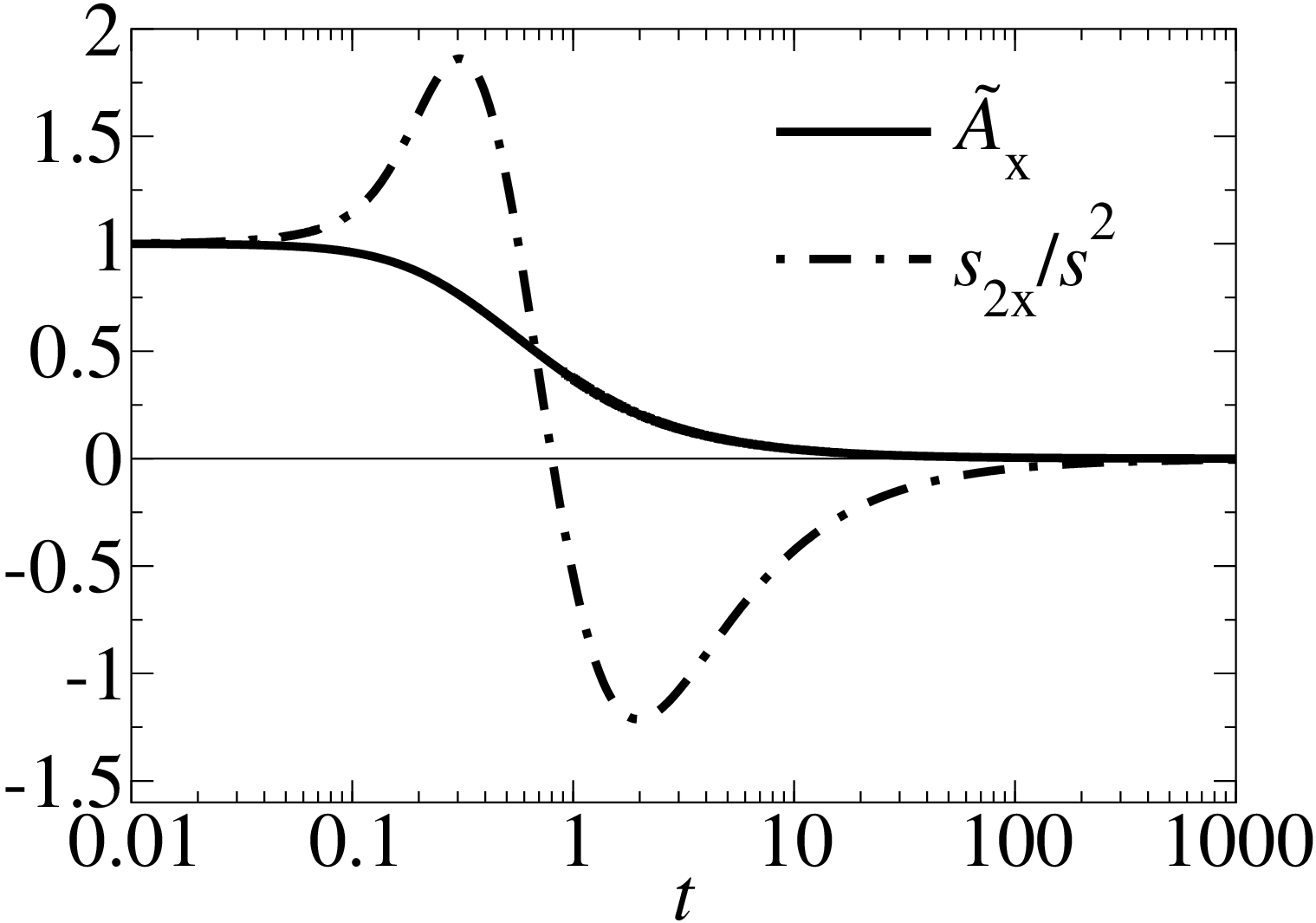}
\includegraphics*[angle=-00,height=2.8cm]{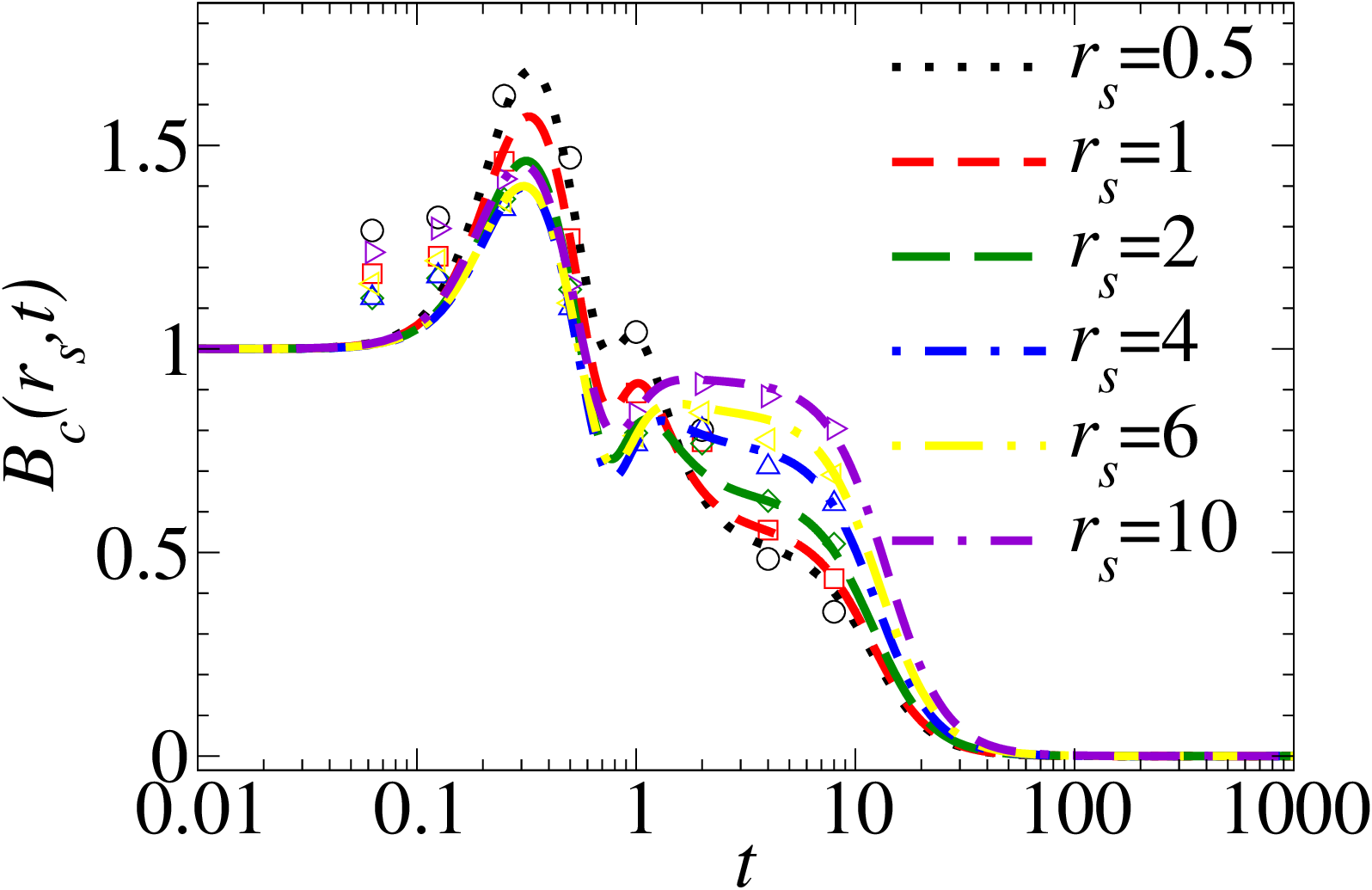}
\caption{
Left: Behavior of $\widetilde A_{\mathrm{x}}$ and 
$s_{\mathrm{2x}}/s^2\equiv \widetilde B_{\mathrm{x}}/\widetilde A_{\mathrm{x}}$
as functions of t. 
Right: Comparison of 
$\widetilde B_{\mathrm{c}}(r_{\mathrm{s}},t)$ 
reference data (symbols) and 
analytical fit (curves) for selected  $r_{\mathrm{s}}$ values.
}
\label{chi-tildeB-tildeBc}
\end{figure}

The left-hand panel of Fig.\ \ref{chi-tildeB-tildeBc} shows both $%
\widetilde A_{\mathrm{x}}$ and the ratio $s_{2\mathrm{x}}/s^2\equiv
\widetilde B_{\mathrm{x}}/\widetilde A_{\mathrm{x}}$ as functions of $t$. $%
\widetilde A_{\mathrm{x}}$ vanishes in the high-T limit, but $\widetilde B_{%
\mathrm{x}}$ decays more rapidly (see Ref.\ \cite{FDfits} for the relevant
asymptotic expansions) such that the ratio $\widetilde B_{\mathrm{x}%
}/\widetilde A_{\mathrm{x}}$ eventually vanishes as well. That guarantees
satisfaction of the correct high-T limit for X (provided that $F_{\mathrm{x}%
}(0)=1$; see additional comments below Eq.\ (\ref{FxVVK})). Further, the
definition Eq.\ (\ref{FxGGA}) guarantees that the X free-energy scales
correctly \cite{Pittalis11,DuftyTrickey16}, ${\mathcal{F}}_{\mathrm{x}}^{%
\mathrm{GGA}}[n_{\lambda},\mathrm{T}]= \lambda{\mathcal{F}}_{\mathrm{x}}^{%
\mathrm{GGA}}[n,\mathrm{T}/\lambda^2]$, with $n_{\lambda}(\mathbf{r}%
)=\lambda^3 n(\lambda \mathbf{r})$.

Because of  Eq.\ (\ref{limschi}), the simplest
approximation for a finite-T X enhancement factor $F_{\mathrm{x}}(s_{2%
\mathrm{x}})$ might seem to be a zero-T GGA X enhancement factor.
That would meet requisite (c) above. However, the distinctive
sign change of $s_{2\mathrm{x}}$ near $t=1$, Fig.\ \ref{chi-tildeB-tildeBc},
precludes straightforward adoption of popular choices such as the PBE
enhancement factor \cite{PerdewBurkeErnzerhof96} because unphysical
poles could result. A more refined finite-T generalization is
required.

A well-behaved X enhancement factor arises from imposition of the
following constraints: %
[i] $F_{\mathrm{x}}(0)=1$ to recover the HEG limit at all T; %
[ii] recovery of the T-dependence of the GE2 
in the small-$s$ limit 
$F_{\mathrm{x}}(s_{2\mathrm{x}}) \approx 1+\nu_{\mathrm{x}} s_{2\mathrm{x}}$
with $\nu_{\mathrm{x}}$ a constant consistent with the $s^2$ coefficient in
the T=0K limit GGA; %
[iii] local satisfaction of the zero-T Lieb-Oxford bound \cite%
{Lieb.Oxford.1981} by requiring $F_{\mathrm{x}}(s_{2\mathrm{x}})\le F_{%
\mathrm{x,max}}=1.804$ (see \cite{PerdewBurkeErnzerhof96}); %
[iv] smooth, non-negative behavior for all $s_{2\mathrm{x}}\in \,
]-\infty,+\infty[$, to match the behavior of exact X at finite-T \cite%
{Greiner.Carrier.Goerling.FT-EXX}.

A simple enhancement factor 
\begin{equation}
F_{\mathrm{x}}(s_{2\mathrm{x}})= 1+\frac{\nu_{\mathrm{x}} s_{2\mathrm{x}}}{%
1+\alpha |s_{2\mathrm{x}}|} \,,  \label{FxVVK}
\end{equation}
with $\alpha=\nu_{\mathrm{x}}/(F_{\mathrm{x,max}}-1)$ satisfies all of those
constraints. Additionally, (\ref{FxVVK}) with suitably chosen constants
recovers PBE X in the zero-T limit: $s_{2\mathrm{x}} \rightarrow s^2$ $%
\Rightarrow$ $\lim_{\mathrm{T}\rightarrow 0}F_{\mathrm{x}}(s_{2\mathrm{x}%
})=1+\nu_{\mathrm{x}}s^2/(1+\alpha s^2)$. 
In the high-T limit, the density-gradient dependence of $s_{\mathrm{2x}}$ is
suppressed by the decaying tail of the $\widetilde B_{\mathrm{x}%
}(t)/\widetilde A_{\mathrm{x}}(t)$ function (see Fig.\ \ref%
{chi-tildeB-tildeBc}), such that $\lim_{\mathrm{T}\rightarrow\infty}F_{%
\mathrm{x}}(s_{2\mathrm{x}})=1$. Constraint [i] thus is satisfied not only
for the strictly homogeneous case ($s=0$), but also for non-uniform
densities with any finite $s$ value. This property is inherited correctly by
the finite-T GGA, Eq.\ (\ref{FxVVK}), from the GE2, Eq.\ (\ref{fxSGA2}).

\textit{Finite-T GGA Correlation}.
Recall that C 
and X  differ in that $\widetilde B_{\mathrm{c}}$ depends upon 
both $n$ and $t$.  The Supplemental Material \cite{SuppMat} 
gives details of the $\widetilde B_{\mathrm{c}}$ analytical fit 
%
developed in this work.  It uses  numerical
results from  Ref.\ \cite{Sjostrom.Daligault.2014A}, static local 
field corrections  \cite%
{Niklasson..Singwi.1975,Gupta.Singwi.1977}, and quantum Monte-Carlo data for
the finite-T HEG \cite{BrownEtAl2013}.
The right-hand panel of Fig.\ \ref{chi-tildeB-tildeBc} shows the 
smooth T and $r_{\mathrm{s}}$ dependencies of $\widetilde B_{\mathrm{c}}$. 
It is everywhere positive, goes to unity
in the zero-$t$ limit (by construction), and vanishes in the high-T limit
(thereby guaranteeing the correct high-T limit for
correlation).
At and below $t \approx 0.2$, enforcement of total entropy positivity
for physical systems necessitated that the $\widetilde B_{\mathrm{c}}$ fit lie
below the data \cite{SuppMat}.

With $\widetilde B_{\mathrm{c}}$ in hand, the fact that the C term in Eq.\ (%
\ref{gxc2}) is proportional to 
$n^{1/3}s^2\widetilde B_{\mathrm{c}}(r_{\mathrm{s}},t) \propto q^2\widetilde
B_{\mathrm{c}}(r_{\mathrm{s}},t)$ (with $q(n,\nabla n)=|\nabla n|/2k_{%
\mathrm{s}}n$ the ground-state variable and $k_{%
\mathrm{s}}=2(3n/\pi)^{1/6}$) 
motivates
definition of the T-dependent reduced density gradient for C as 
\begin{equation}
q_{\mathrm{c}}(n,\nabla n,{\mathrm{T}})=q(n,\nabla n) \sqrt{\widetilde B_{%
\mathrm{c}}(r_{\mathrm{s}},t)} \,,  \label{tc}
\end{equation}

In terms of $q_{\mathrm{c}}$, the finite-T GGA C functional is
determined by imposition of the following conditions. The functional must %
[v] provide the correct HEG limit both at zero- and at finite-T,
i.e.\ reduce to the LDA C (free-) energy; %
[vi] reproduce the slowly varying regime correctly; for T$>0$ the correct
T-dependence in that regime is given by $\widetilde B_{\mathrm{c}}$; %
[vii] satisfy known T=0K constraints for C (e.g.\ Ref.\ \cite%
{PerdewBurkeErnzerhof96}); %
[viii] reduce to the LDA C free energy in the high-T limit for any $n$ with
finite reduced gradient $q$ (in consequence of the finite-T gradient
expansion).

The simplest approximation which satisfies all these constraints is based on
a known zero-T GGA correlation functional (which satisfies [vii]).  
T-dependence is introduced by adapting the PBE form of C energy per
particle (spin-unpolarized) to become 
\begin{equation}
\!\!\!\!\! f_{\mathrm{c}}^{\mathrm{GGA}}(n,\nabla n,{\mathrm{T}})= f_{\mathrm{c}}^{%
\mathrm{LDA}}(n,{\mathrm{T}}) +H\Big(f_{\mathrm{c}}^{\mathrm{LDA}%
},\zeta=0,q_{\mathrm{c}}\Big) \,,  \label{fcGGA}
\end{equation}
where $f_{\mathrm{c}}^{\mathrm{LDA}}$ is the LDA correlation free-energy per
particle, $\zeta$ is the spin polarization fraction, and $H$ is as defined
in PBE \cite{PerdewBurkeErnzerhof96}, but with substitutions as shown. 
Details are in Ref.\ \cite{SuppMat}. Thus the GGA
correlation free-energy is ${\mathcal{F}}_{\mathrm{c}}^{\mathrm{GGA}}[n,%
\mathrm{T}]=\int nf_{\mathrm{c}}^{\mathrm{GGA}}(n,\nabla n,{\mathrm{T}})d%
\mathbf{r}$. %
In the rapidly varying case, 
$f_{\mathrm{c}}^{\mathrm{GGA}}$ vanishes due to 
a ground state PBE C functional property, 
$\lim_{q_{\mathrm{c}}\rightarrow \infty}H=-f_{\mathrm{c}}^{\mathrm{LDA}}$. 
In the slowly varying regime, 
$f_{\mathrm{c}}^{\mathrm{GGA}}$ 
recovers the second-order gradient expansion (see Refs.\ \cite%
{PerdewBurkeErnzerhof96,Wang.Perdew.1991}) with T-dependence described by $%
\widetilde B_{\mathrm{c}}$. Together with the GE2 for X, that also
provides the correct XC T-dependence defined by Eqs.\ (\ref{gxc})-(\ref{gxc2}%
).

Because $\widetilde B_{\mathrm{c}}$ vanishes in the high-T limit, 
requirement [viii] is satisfied (analogously with the X case) for all
densities with finite values of the variable $q$. This follows from 
$\lim_{\mathrm{T}\rightarrow\infty} H\Big(f_{\mathrm{c}}^{\mathrm{LDA}%
},\zeta,q_{\mathrm{c}}\Big)=0$. Thus the GGA XC free-energy functional, ${%
\mathcal{F}}_{\mathrm{xc}}^{\mathrm{GGA}}[n,\mathrm{T}]\equiv {\mathcal{F}}_{%
\mathrm{x}}^{\mathrm{GGA}}[n,\mathrm{T}] +{\mathcal{F}}_{\mathrm{c}}^{%
\mathrm{GGA}}[n,\mathrm{T}]$, reduces in the high-T limit to the LDA XC
free-energy 
and eventually vanishes, 
\begin{equation}
\lim_{\mathrm{T}\rightarrow \infty}({\mathcal{F}}_{\mathrm{xc}}^{\mathrm{GGA}%
}[n,\mathrm{T}] -{\mathcal{F}}_{\mathrm{xc}}^{\mathrm{LDA}}[n,\mathrm{T}])=0
\, ,  \label{FxcGGA-highT}
\end{equation}
for any density $n$ with finite reduced gradients $s$ and $q$. 

We used PBE values, $\nu_{\mathrm{x}}=0.21951$ in Eq.\ (\ref{FxVVK}), and $%
\beta_{\mathrm{c}}=0.066725$ in $H$, Eq.\ (\ref{fcGGA}). (Remark: to avoid
notational ambiguity, $\nu_{\mathrm{x}}$ and $\beta_{\mathrm{c}}$ 
are the constants denoted $\mu$ and $\beta$ in Ref.\ \cite%
{PerdewBurkeErnzerhof96}.) 
Thus the T=0K limit of our functional, ${\mathcal{F}}_{\mathrm{xc}}^{\mathrm{GGA}}[n,\mathrm{T}]$, is the ground-state PBE functional, with the minor difference 
that we use the corrected KSDT (corrKSDT) parametrization
as a suitably accurate LDA XC free-energy expression 
\cite{KSDT,corrKSDT}. 
Note, however, that virtually any ground-state GGA XC functional
can be extended systematically into an XC free energy functional 
by use of the framework presented above.

\begin{figure}
\includegraphics*[angle=-00,height=2.82cm]{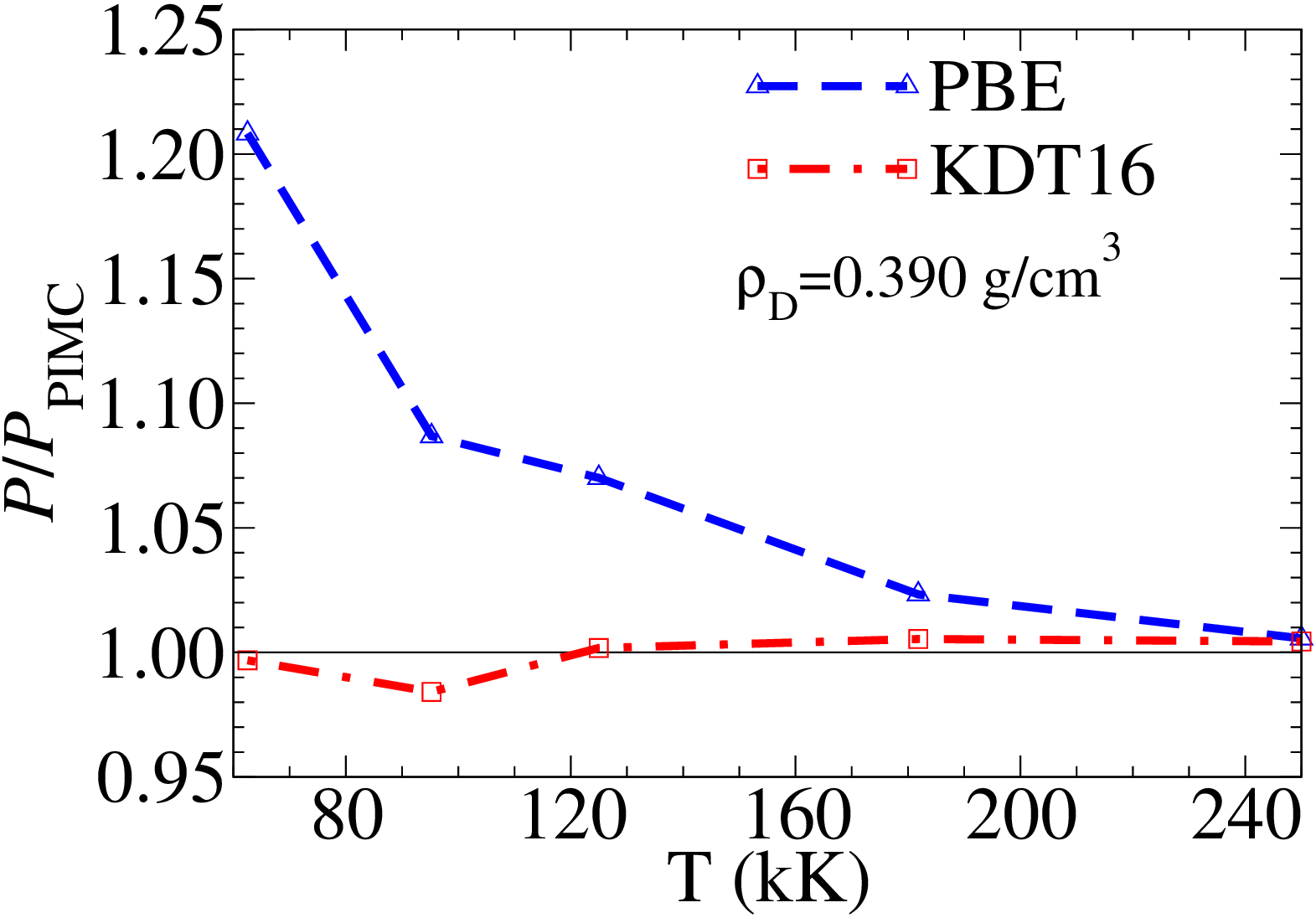}
\includegraphics*[angle=-00,height=2.82cm]{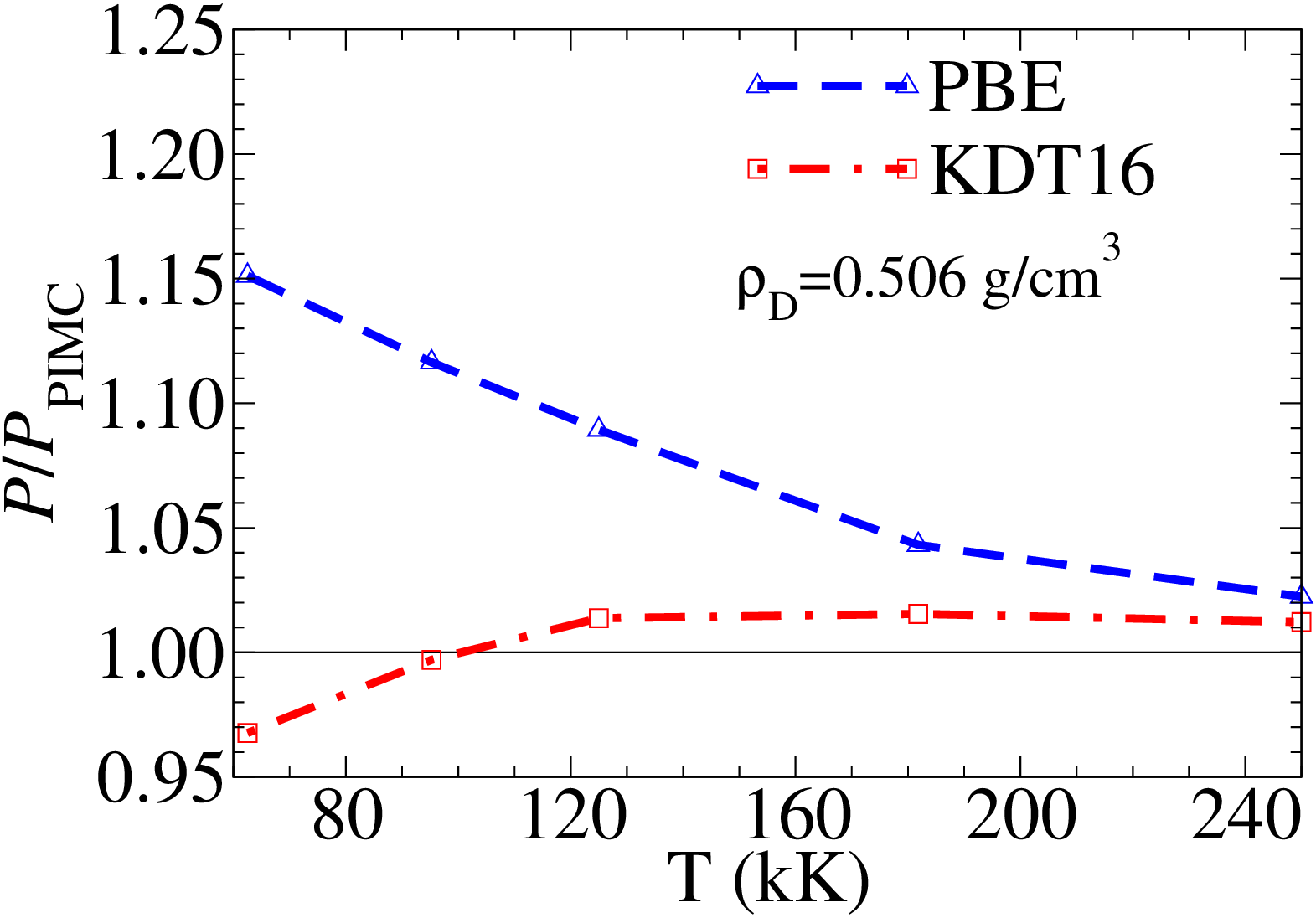}
\caption{
Ratios of Deuterium electronic pressure versus  
T for the free-energy GGA (``KDT16'') and ground-state PBE  
XC functionals, to PIMC reference results.
AIMD PAW 
simulations, $\Gamma$-point only, 
for 64 atoms 
(4500 steps, T$\le 125$kK) and 32 atoms (4500 steps, T$\ge 125$kK); timestep 25-50 asec. 
}
\label{Pinst-D128-D64}
\end{figure}

\textit{Exemplary WDM Applications}
An AIMD simulation which directly probes the accuracy of the new GGA 
functional (``KDT16''), is for deuterium at two 
material densities and T well into the WDM regime. 
Figure \ref{Pinst-D128-D64} compares 
KDT16 and PBE pressures $P$  
relative to high-quality PIMC data \cite{Hu.Militzer..2011}  
(shown for intermediate and high T  only 
due to low-T PIMC limitations \cite{FxcSignif}). 
PBE 
{\it systematically} over-estimates the pressure. 
The deviation is significant at T=62.5, 95.25 and 125kK for both 
material densities, 
then decreases 
as the 
non-interacting free-energy dominates in the high-T limit.
In contrast, the KDT16 pressures are in 
excellent agreement with the  PIMC values for 
the entire T-range, with relative deviations $\le$ 3\%. 

This is a crucial finding two ways.  First is that PIMC codes
are not widely available, they are expensive to run, and PIMC 
itself is limited as to how far down in T
it can go.  Second is that hydrodynamic simulations of cryogenic
inertial confinement implosions using the PIMC equation of state (EOS)
tables found significant differences with respect to simulations based
on the SESAME tables \cite{HuMilitzerGoncharovSkupsky2010}.  The
percentage shifts of pressures from KDT16 versus PBE are comparable 
to the PIMC to SESAME $P$ shifts, so using KDT16 instead of PBE
should have similar impact on hydrodynamic simulations. Note 
consistency with the  
sensitivity of the deuterium principal Hugoniot to ground-state 
XC functional details \cite{KnudsonDesjarlais17}.

\begin{figure}[tbp]
\includegraphics*[angle=-00,height=2.82cm]{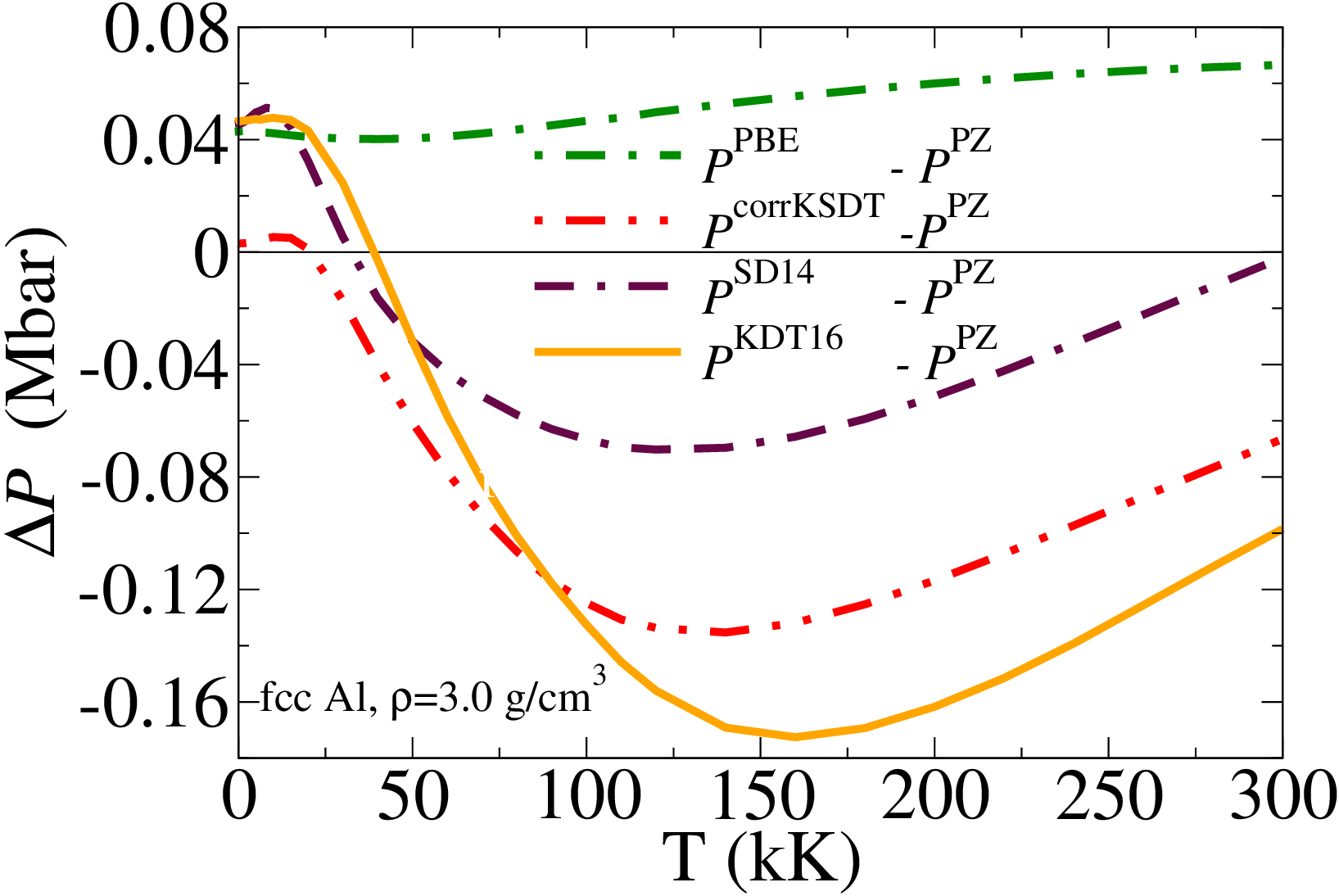}
\includegraphics*[angle=-00,height=2.82cm]{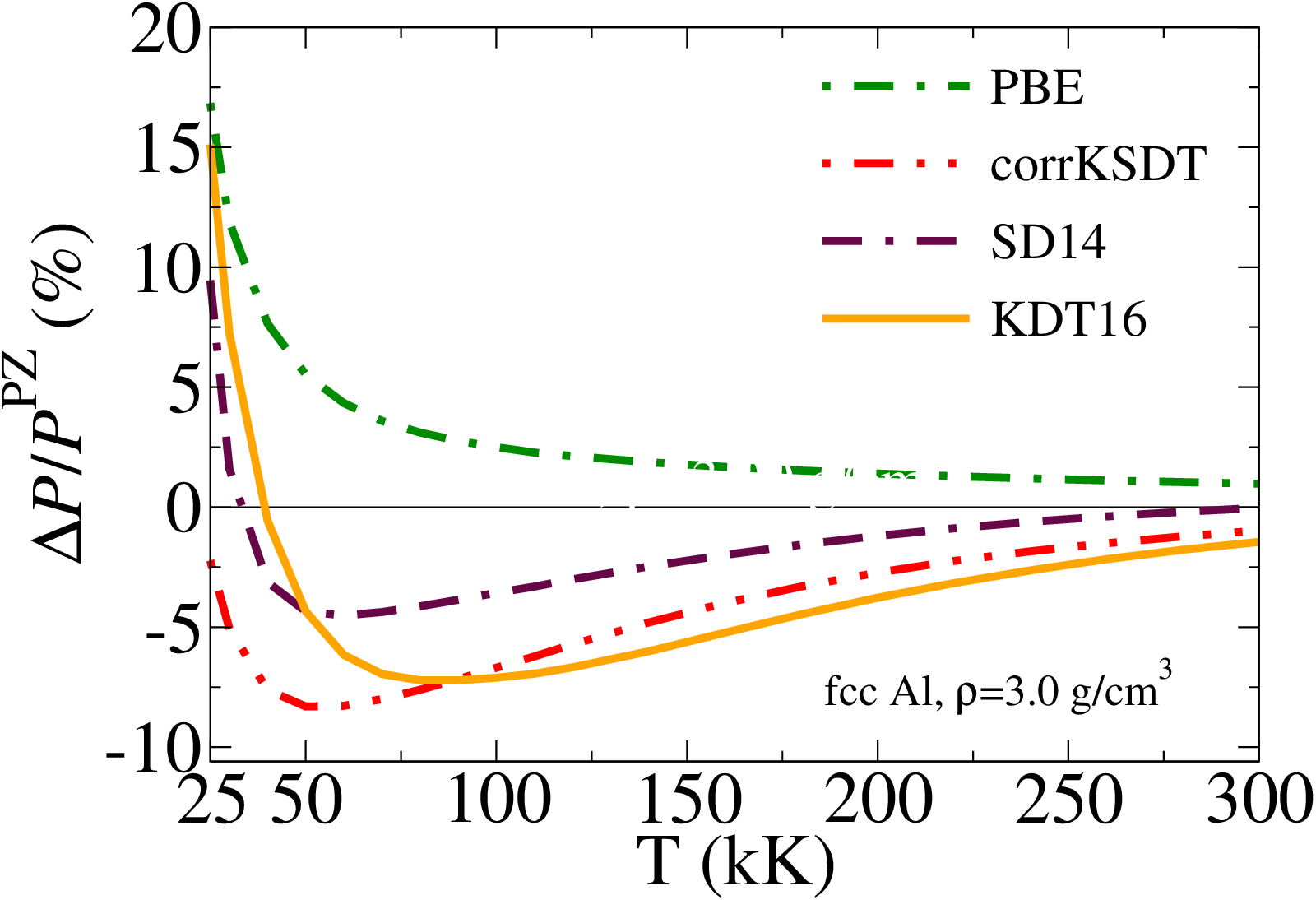}
\caption{Left: Electronic pressure differences $\Delta P({\mathrm T})$ for 
the new KDT16 GGA, SD14 mixed LDA-GGA, corrKSDT LDA, and ground-state PBE XC
functionals, all referenced to PZ ground-state LDA values. Fcc Al, 
3.0 g/cm$^3$. Right: Relative differences.
}
\label{DeltaPSD14vsPresent}
\end{figure}

To isolate static lattice effects, the equation of state
for fcc Al
over a wide T range, $0 \le {\mathrm{T}}\le 300$kK, at slightly
compressed 
material density $\rho=3.0$ g/cm$^3$ (as 
used in LCLS experiments \cite{LCLS.2015}) was calculated from three 
thermal XC functionals, the
new KDT16, SD14, corrKSDT, and two ground-state functionals 
(PZ \cite{PZ81} LDA and PBE GGA). Because LDA is widely viewed as 
good for metals,  PZ is used as the reference.
 Calculations were  
done with a locally modified version of Quantum-Espresso \cite{QE,ProfessQE}.
For the PZ and corrKSDT functionals we used a 
PAW data set built with PZ XC. For KDT16, SD14, and PBE
we used the PBE PAW data set. All the calculations
were otherwise self-consistent.

The resulting pressure differences shown in Fig.\ \ref{DeltaPSD14vsPresent} distinguish 
XC inhomogeneity effects (see $(P^{\mathrm{PBE}}-P^{\mathrm{PZ}})$),
thermal XC effects at the LDA level of refinement (see $(P^{\mathrm{corrKSDT}}-P^{%
\mathrm{PZ}})$), and the combined XC effects at the GGA refinement level in
($P^{\mathrm{KDT16}}-P^{\mathrm{PZ}}$). The new KDT16 functional 
interpolates smoothly between the PBE values at low-T and the corrKSDT (LDA) 
values at 
high-T (not fully shown). Pressure differences from both KDT16 
and from corrKSDT 
have their maximum magnitude at intermediate-T, then
decrease. Crucially, the ground state approximation, i.e.\ ${\mathcal{F}}_{%
\mathrm{xc}}^{\mathrm{GGA}}[n,\mathrm{T}] \approx E_{\mathrm{xc}}^{\mathrm{GGA}}[n]$ 
systematically overestimates the pressure by as much as $%
\approx$10\% at T between $\approx 40$ and $100$kK. 
The pressures from all proper functionals eventually go to a common 
high-T limit as 
the XC contribution becomes negligible compared to the non-interacting
free-energy. However, the behavior en route to that 
limit is qualitatively different for a free-energy GGA 
versus a ground-state functional.  Note that SD14 pressures start to 
deviate significantly
from the values given by KDT16 at ${\mathrm{T}}=50$kK ($t \approx 0.27$), 
roughly the beginning
of the WDM regime.


Most importantly, the $\Delta P({\mathrm T})$ behavior of \emph{all} three
T-dependent XC functionals differs \emph{qualitatively} from that from  
PBE.  $\Delta P({\mathrm T})$ for PBE is uniformly
positive, whereas $\Delta P(T)$ is negative for each of the 
explicitly T-dependent functionals above some relatively small T. 
Almost surely, therefore, the T-dependence from PBE is wrong.  This
qualitative distinction in the calculated EOS will have direct 
consequences for material predictions. An example is calibration 
of effective potential approaches.  
Ref.\ \cite{WunschVorbergerGericke2009} compared the Al ion-pair
distribution from such a scheme with PBE AIMD
results for T=1.1 eV, $\rho =$ 3.4 g/cm$^3$ and deemed the 
agreement satisfactory.  A similar comparison for the so-called
ion feature recently is in Ref.\ 
\cite{Harbour..Lewis2016}.  For T=10 eV and $\rho$ = 8.1 g/cm$^3$, 
their ``NPA'' model with T$_{ion}$ = 1.8 eV compares much
more favorably with the experimental peak height than the AIMD result
with PBE XC.  Indeed,  R\"uter and Redmer \cite{RuterRedmer2014}
had done an AIMD PBE static structure factor calculation 
at T=1,023 K ($\approx$ 0.1 eV), $\rho$ = 2.35 g/cm$^3$ that agreed 
well with experiment for Al. But at T = 10 eV, $\rho$= 8.1
g/cm$^3$ their AIMD-PBE calculation seriously 
underestimated the experimental ion-feature peak height.
They attributed this discrepancy 
to omission of core electron effects but had no way to 
assess EOS effects which we have shown here to be substantial.

\begin{figure}
\hspace{-0.15cm}
\includegraphics*[angle=-00,width=6.20cm]{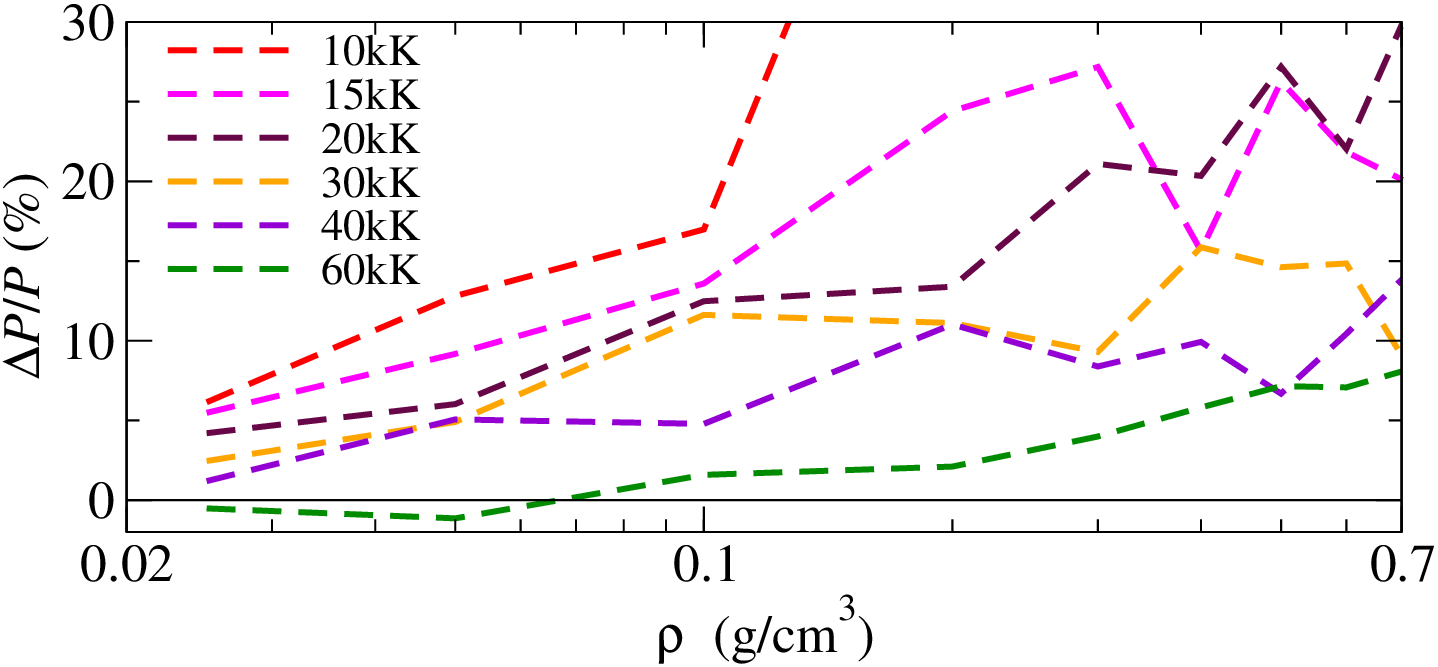}
\caption{
Relative difference in Al total pressure along six
isotherms (10, 15, 20, 30, 40, 60 kK) for KDT16 and PBE XC functionals
plotted as $(P^{\mathrm{PBE}}-P^{\mathrm{KDT16}})/P^{\mathrm{KDT16}} \times 100$. 
}
\label{Al-EOS}
\vspace*{-0.1cm}
\end{figure}

Given the importance of EOS shifts, our final example is 
low density Al; its measured  
electrical conductivity 
exhibits pronounced system  
density dependence \cite{DeSilva.Katsouros.1998}.  Fig.\ \ref{Al-EOS} shows 
that the low-density Al total pressure is strongly affected by 
use of the fully gradient-dependent and T-dependent XC functional. 
Shifts relative to it caused by the ground-state PBE approximation range as
high as  $\approx$50\% (T=10kK) down to 5\% (T=60kK).
Clearly  
there is no simple rule-of-thumb correction for the ground-state functional
data. Nor, on fundamental grounds, is there any reason to assume
that it is the better of the two functionals.  (Remark: Figure S2 in the 
Supplemental Material shows the ineffectiveness in identifying errors
via direct comparison of KDT16 and PBE results.) 

\textit{Implications and Summary}. The non-empirical KDT16 GGA
XC free-energy functional 
is more systematically constructed and general 
than the only previous attempt at a finite-T GGA 
\cite{Sjostrom.Daligault.2014A}. KDT16 treats both 
density inhomogeneity 
and T-dependence effects yet distinguishes them clearly.
Three rather different example calculations 
show its accuracy and value.
They also confirm that   
ground-state GGA functionals are not routinely reliable as
free energy functionals 
\cite{FxcSignif,DanelKazandjianPrion2016,Schoof..Bonitz.PRL2015,Dornheim..Bonitz.PRL2016}.

The new ${\mathcal{F}}_{%
\mathrm{xc}}^{\mathrm{GGA}}$ has no empirical parameters. As with 
ground-state functionals, it involves  design choice \cite%
{PerdewBurkeErnzerhof96,PBEsol,PBEmol} for the gradient-expansion
coefficient for X ($\nu_{\mathrm{x}}$) and the related C parameter  ($%
\beta_{\mathrm{c}}$).  Yet the underlying procedure is general.  
Analysis of the XC gradient expansion leads to 
 appropriate T-dependent variables for X and C. Together 
with the new $\widetilde B_{\mathrm{c}}(r_{\mathrm{s}},t)$  parametrization 
and the LDA free-energy parametrization \cite{KSDT,corrKSDT}, 
one has the basis for GGA XC free-energy
functional development.  Virtually any ground-state GGA XC functional
thus can be extended 
systematically into an XC free energy functional by use of the T-dependent
variables Eqs.\ (\ref{sx}), (\ref{tc}) within this framework.

\begin{acknowledgments}
We thank Travis Sjostrom for providing the
numerical $ g_{\mathrm{xc}}^{(2)}(n,\mathrm{T})$ data 
and for helpful comments on 
an earlier version of the manuscript. We thank Kieron Burke, Florian
Eich, and Giovanni Vignale for instructive comments.  We thank the
Univ.\ of Florida Research Computing organization for
computational resources and technical support. This work was supported
by the U.S.\ Dept.\ of Energy BES grant DE-SC0002139.
VVK also acknowledges support by the Dept. of Energy National 
Nuclear Security Administration under Award Number DE-NA0001944 at the final 
stage of work.
\end{acknowledgments}


\begin{thebibliography}{99}
\bibitem{CoeffComment2} Whether the coefficient in Eq.\ (\ref{fxSGA2}) is 8/81
or 10/81 as in Refs.\ \cite{FactorRefs} is irrelevant. Here  
we must match the small $s^2$ behavior of the ground state GGA X functional
chosen for the T$\rightarrow 0$K limit, i.e., PBE.
\end{thebibliography}
\end{document}